\documentclass[%
pre,
 amsmath,amssymb,
 reprint,%
]{revtex4-1}

\usepackage{graphicx}
\usepackage{dcolumn}
\usepackage{bm}

\usepackage[utf8]{inputenc}
\usepackage[T1]{fontenc}
\usepackage{mathptmx}
\usepackage{commath}

\DeclareRobustCommand{\uvec}[1]{{%
  \ifcsname uvec#1\endcsname
     \csname uvec#1\endcsname
   \else
    \bm{\hat{\mathbf{#1}}}%
   \fi
}}

\providecommand{\keywords}[1]
{
  \small	
  \textbf{\textit{Keywords---}} #1
}

\usepackage{url}
\usepackage{hyperref}

\begin{document}

\preprint{AIP/123-QED}

\title[]{Bayesian optimization of discrete dislocation plasticity\\ of two-dimensional precipitation hardened crystals}

\author{Mika Sarvilahti}
\email{mika.sarvilahti@tuni.fi}
\author{Lasse Laurson}
\affiliation{ 
Computational Physics Laboratory, Tampere University, \\
P.O. Box 692, FI-33014 Tampere, Finland
}%

\date{\today}

\begin{abstract}
\noindent
Discovering relationships between materials’ microstructures and mechanical properties is a key goal of materials science. Here, we outline a strategy exploiting Bayesian optimization to efficiently search the multidimensional space of microstructures, defined here by the size distribution of precipitates (fixed impurities or inclusions acting as obstacles for dislocation motion) within a simple two-dimensional discrete dislocation dynamics model. The aim is to design a microstructure optimizing a given mechanical property, e.g., maximizing the expected value of shear stress for a given strain. The problem of finding the optimal discretized shape for a distribution involves a norm constraint, and we find that sampling the space of possible solutions should be done in a specific way in order to avoid convergence problems. To this end, we propose a general mathematical approach that can be used to generate trial solutions uniformly at random while enforcing an Euclidean norm constraint. Both equality and inequality constraints are considered. A simple technique can then be used to convert between Euclidean and other Lebesgue $p$-norm (the 1-norm in particular) constrained representations. Considering different dislocation-precipitate interaction potentials, we demonstrate the convergence of the algorithm to the optimal solution and discuss its possible extensions to the more complex and realistic case of three-dimensional dislocation systems where also the optimization of precipitate shapes could be considered.
\end{abstract}

\keywords{crystalline materials; plastic deformation; dislocation systems; precipitates; constrained optimization}

\maketitle



\section{Introduction}

\noindent
The need to develop novel materials with desired properties for applications is the driving force behind much of materials science. One way of framing the problem is in terms of structure-property relationships~\cite{ghosh2011computational}, where one aims at establishing relations between, say, the microstructural features of a solid material and its mechanical properties~\cite{raabe2014alloy}. In general, the problem is very challenging due to the combination of high dimensionality of microstructural descriptors (due to the very large number of different microstructural features~\cite{staron2017neutrons}) and non-linearities and statistical fluctuations in the material response to external stimuli~\cite{zaiser2006scale, uchic2009plasticity}. For these reasons, conventional materials design strategies relying essentially on a combination of educated guesses and trial and error are sub-optimal and constitute a bottleneck for discovery of novel materials.

Recent years have witnessed the emergence of “smart” methods in the toolbox of materials scientists, such as machine learning (ML) and optimization algorithms, which are used to discover previously unknown dependencies of material properties on a wide range of microstructural parameters~\cite{pilania2021machine, karniadakis2021physics}. Indeed, such developments are currently spawning a new research field sometimes referred to as {\it materials informatics}~\cite{agrawal2016perspective}. However, a large fraction of applications of “ML for materials” are currently limited to considering atomistic and molecular properties of materials in fields such as quantum chemistry~\cite{von2018quantum}. Yet, the macroscopic mechanical properties of realistic crystalline materials are largely controlled by the microstructural features on scales well above the atomic/molecular scale, calling for novel applications of ML to discover novel microstructure-property relations using microstructural features on a coarse-grained scale as input. 

Here, we present an approach based on Bayesian optimization to find optimal values for the properties, such as the size distribution, of precipitate particles within crystals resulting in desired mechanical properties, such as maximal stress at a given strain. Bayesian optimization~\cite{pelikan1999bayesian, mockus2012bayesian, frazier2018tutorial} is known to be an excellent choice for optimization problems such as the present one in which evaluating a data point (here, performing a discrete dislocation dynamics (DDD) simulation) is computationally expensive and the outcome is stochastic (noisy). The method can be applied to various problems, and there has already been success in the context of optimization of, e.g., atomistic structures~\cite{todorovic2019bayesian} and metamaterials~\cite{ vangelatos2021strength}. Bayesian optimization has also been used to calibrate the parameters of a gradient plasticity model~\cite{tan2021} that predicts the plastic size effects of micropillars.

In this work, we apply Bayesian optimization to 2D dislocation systems. The stress-strain response of such pure dislocation systems has been studied both using stress-controlled~\cite{ispanovity2014avalanches, szabo2015} and strain-controlled~\cite{miguel2002dislocation, kurunczipapp2021dislocation} loading, along with recent attempts to predict the response to external stresses by using machine learning techniques~\cite{salmenjoki2018machine, minkowski2022machine}. Like in these works, we impose periodic boundary conditions for simplicity, although it should be mentioned that recently, various methods have been developed for simulating finite systems~\cite{pan2017, berta2020, shima2022}. Our system also contains fixed round precipitates (or solute clusters) acting as obstacles for the moving dislocations. Precipitation is known to cause various pinning effects, which have been studied with 2D DDD simulations in the case of identical precipitates or pinning centres~\cite{ovaska2015quenched}. Here, we go further and allow the precipitates to be of varying sizes with the aim of employing Bayesian optimization to find the optimal shape for a discretized size distribution, subject to the constraint of a fixed volume fraction (area fraction in 2D) of precipitates, resulting in a designed mechanical response of the material. The design objective in our case is to maximize the average shear stress required to produce a certain value of strain.

An alternative 2D dislocation modelling choice would be to represent dislocations as lines on a single glide plane~\cite{mohles1999simulation}, leading to a more complicated dislocation-precipitation interaction, but this would also make the set of dislocations form an effectively one-dimensional pileup system~\cite{sarvilahti2020machine,moretti2004depinning}, which is unable to capture some of the phenomena happening in systems with multiple glide planes, typically related to the competition between dislocation jamming~\cite{miguel2002dislocation} and pinning due to obstacles~\cite{ovaska2015quenched}. Certain cross-section models called the 2.5D models~\cite{keralavarma2016} have also been developed with the aim of incorporating some of these effects into 2D simulations by utilizing statistics related to such phenomena collected from 3D simulations.

Realistic but computationally more demanding 3D DDD models have also been considered, both without~\cite{lehtinen2016glassy} and with precipitates~\cite{salmenjoki2020plastic, liu2021dislocation}. We intend to extend our optimization study to such systems after the approach has been tested and polished for the 2D case, which is the focus of this work. In the 3D case, bypassing mechanisms between dislocations and solute clusters have been observed to change considerably depending on the cluster size~\cite{fang2019statistical, eswarappa2022strengthening}, which motivates our attempts to find ways of taking advantage of such mechanisms when designing materials.

Our work starts by explaining the specifics of the 2D DDD model to be studied in Section~\ref{sec:model}. Section~\ref{sec:bayes} introduces the Bayesian optimization method. In Section~\ref{sec:generation}, we investigate the specific problem of generating feasible points, which turns out to be a critical point for ensuring optimization convergence. The proof-of-concept test results are presented in Section~\ref{sec:results}, followed by discussion and conclusions in Section~\ref{sec:discussion}.

\section{The 2D DDD model}
\label{sec:model}

\noindent
The discrete dislocation model in two dimensions starts with $N=64$ dislocation points, representing cross-sections of edge dislocation lines, placed on a square simulation box. The dislocations have their Burgers vector along the $x$-axis of the $x y$-plane, and the direction of the Burgers vector is $\pm \uvec{x}$ with even portions of both signs. Each dislocation generates a shear stress field~\cite{ovaska2015quenched, anderson2017theory}
\begin{equation}
\label{eq:disdis}
\sigma_{yx} (\mathbf{r}, \, s b) = \frac{\mu s b}{2 \pi (1-\nu)} \frac{x(x^2-y^2)}{(x^2+y^2)^2},
\end{equation}
where $\mathbf{r}=\begin{bmatrix} x \\ y \end{bmatrix}$ is position with respect to the dislocation with Burger's vector sign $s$ and magnitude $b$, in a material with shear modulus $\mu$ and Poisson's ratio $\nu$. The dislocations are allowed to move only in the $x$-direction.

\begin{figure}[t!]
    \begin{center}
  \includegraphics[width=\columnwidth]{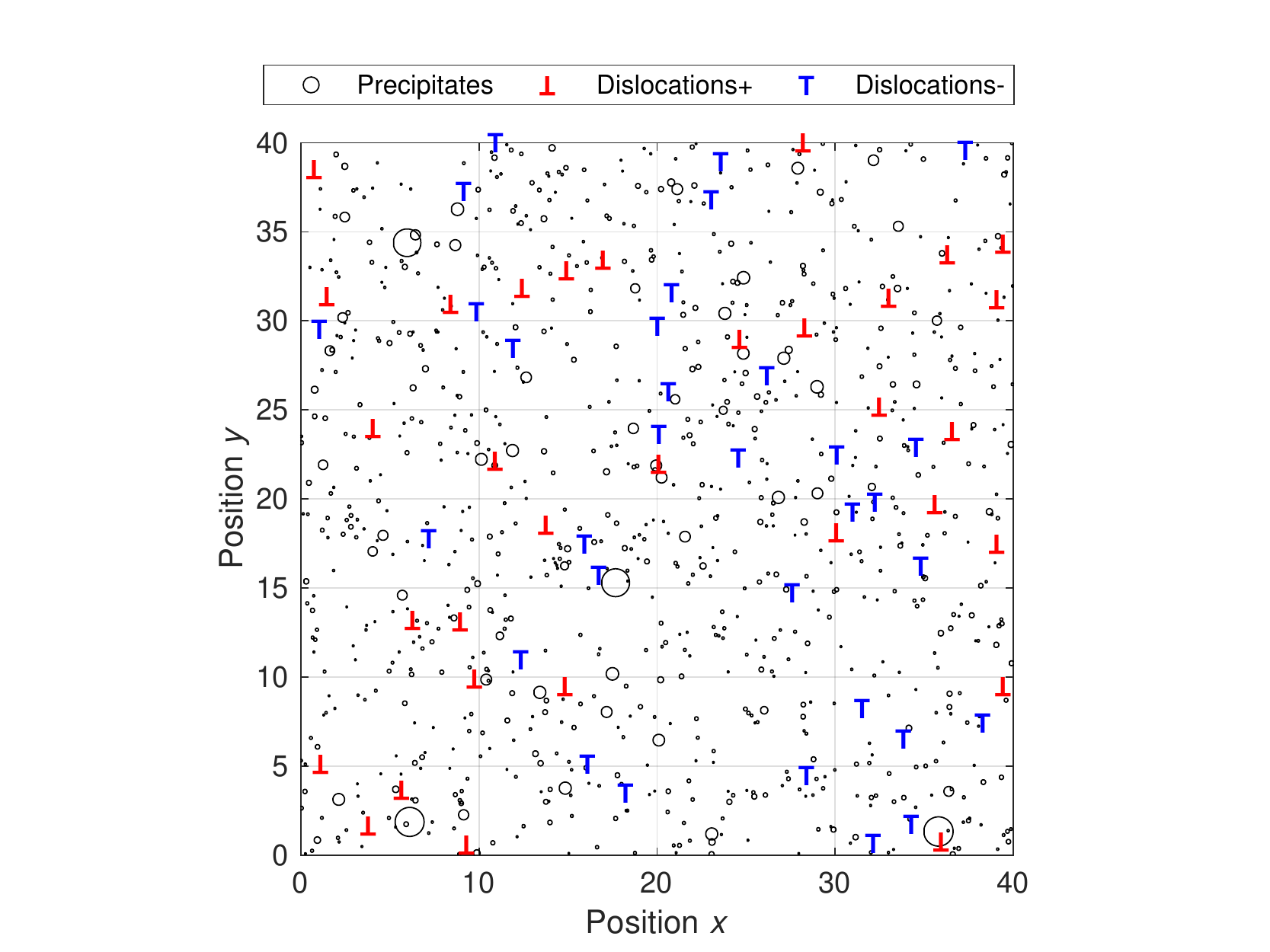}
  \caption{An example of a relaxed dislocation configuration with $N=64$ dislocations and periodic unit cell side length $L=40$. The randomly positioned precipitates are shown as circles, and the T-symbols correspond to dislocations, with different orientation and color for opposite Burgers vectors.}
  \label{fig:b1}
\end{center}
\end{figure}

The system also contains spherical precipitates which occupy 3\% of the total area. An example of a relaxed configuration is presented in Figure~\ref{fig:b1}. The precipitates are fixed in place and interact with dislocations through a spherically symmetric Gaussian potential $U$, 
\begin{equation}
\label{eq:R1potential}
U(r, \, R) = AR\exp{\left(-\frac{1}{2}\left(\frac{r}{R}\right)^2\right)},
\end{equation}
where $r$ is the distance from the center of the precipitate, $R$ is the radius (the size) of the precipitate, and $A=0.5$ is a constant scale parameter. The interaction force is the negative partial derivative of $U$ with respect to the $x$-coordinate.

We also consider an alternative potential $U_{alt}$ having a stronger scaling for the force magnitude with respect to $R$:
\begin{equation}
\label{eq:R2potential}
U_{alt}(r, \, R) = BR^2\exp{\left(-\frac{1}{2}\left(\frac{r}{R}\right)^2\right)},
\end{equation}
where $B=5.0$ is another constant. Changing to this potential changes the expected location of the optimum as the scaling of the interaction force with respect to the precipitate size is stronger.

All dislocations in this model are chosen to have the same Burger's vector magnitude $b$. Then, the equation of motion~\cite{ovaska2015quenched} for a dislocation positioned at $\mathbf{r}_i$ is
\begin{equation}
\label{eq:motion}
\frac{\mathrm{d}}{\mathrm{d}t}x_i = s_i b^2 \chi \left[ \sum_{j \neq i} \sigma_{yx} (\mathbf{r}_i - \mathbf{r}_j, \, s_j b) + \sigma_{ext} \right] - \chi \frac{\partial}{\partial x} U_{total}(\mathbf{r}_i),
\end{equation}
with dislocation mobility $\chi$, external shear stress $\sigma_{ext}$, and $U_{total}(\mathbf{r}) = \sum_k U (||\mathbf{r} - \mathbf{r}_k||_2, \, R_k)$, where $k$ iterates over every precipitate. A simplified unit system is chosen by setting the variables $b$, $\chi$ and $\frac{\mu}{2 \pi (1-\nu)}$ equal to 1. The square-shaped system's side length is $L=40$. Furthermore, we impose periodic boundary conditions and take all the periodic images of dislocations into account in a finite form by modifying~\cite{anderson2017theory} the long-range shear stress field formula of Eq.~(\ref{eq:disdis}). Then, integrating the equation of motion while slowly (quasistatically) increasing the external stress from zero (after relaxing the system without external stress) makes the dislocations move, and the strain~\cite{ovaska2015quenched}
\begin{equation}
\epsilon(t) = \frac{1}{L^2}\sum_{i=1}^{N}s_i b\left[x_i(t)-x_i(0)\right]
\end{equation}
is measured. Two oppositely signed dislocations are annihilated if their distance becomes less than $b$ (= 1 in the chosen unit system). The simulation ends when the average strain $\epsilon$ reaches the value $0.2$ (a typical value in some recent works~\cite{kurunczipapp2021dislocation, salmenjoki2018machine, minkowski2022machine}).

\begin{figure}[t!]
    \begin{center}
  \includegraphics[width=\columnwidth]{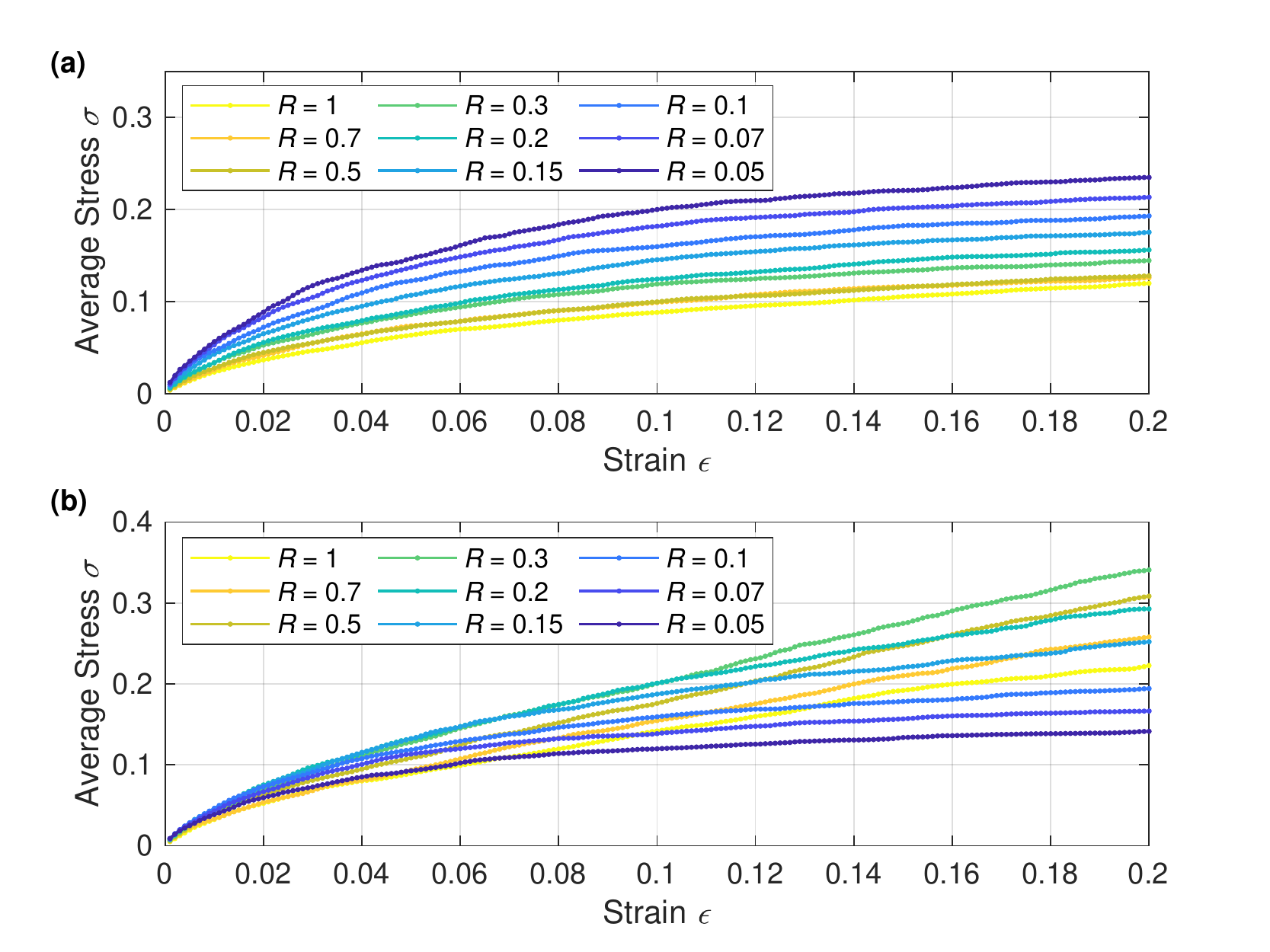}
  \caption{Mean stress--strain curves
    $\sigma (\epsilon)$ (averaged curves over 100 random configurations) for nine different precipitation sizes $R$ (number of dislocations $N=64$, volume fraction of precipitates $=0.03$, $\delta$ distribution for precipitate sizes), using the interaction potential (a) $U$ from Eq.~(\ref{eq:R1potential}) or (b) $U_{alt}$ from Eq.~(\ref{eq:R2potential}). The size that maximizes stress depends on the chosen potential. Our aim in this work is to maximize the average stress needed for strain $\epsilon=0.2$ with respect to a precipitate size distribution that can take any shape (given a finite resolution) instead of assuming one (such as a $\delta$ or a Gaussian shape) for the distribution.}
  \label{fig:b2}
\end{center}
\end{figure}

The response of a dislocation system to external shear stress is described by a stress-strain curve. Figure~\ref{fig:b2} shows how the average stress-strain curve depends on the precipitate size and on the choice of the interaction potential for the case where all precipitates have the same size. In contrast to the average response, an individual response typically alternates between jammed states (with slight elastic deformation) and strain bursts (plastic deformation caused by dislocation avalanches), producing staircase-like stress-strain curves~\cite{ispanovity2014avalanches, salmenjoki2018machine}. Avalanches become more prominent in the response curve the smaller the system is, and there can be significant differences between the responses of systems built from the same recipe but having different random configurations. From the perspective of optimization, this can be viewed as noise. Therefore, information should be collected from multiple configurations when determining the links between recipes and responses.

In practice, the level of noise in the system response can be decreased by taking the sample mean over $M$ random configurations. An alternative way to reduce noise would be to increase the size of the dislocation system. We know that the computational cost of a DDD simulation scales as $\sim N^2$, where $N$ is the number of dislocations, whereas running $M$ multiple simulations (parallel runs possible) simply multiplies the computational cost by $M$, which means linear scaling $\sim N$ with respect to the total number of dislocations over the $M$ systems. Both methods typically decrease noise as $\sim 1/\sqrt{N}$ (may not be exact when increasing the system size~\cite{ispanovity2014avalanches}, but still close to). This suggests that the most efficient way to reduce noise is to run many systems in parallel and to have the averaged results be the observations for the optimization. However, the system size should be large enough so that all relevant physical phenomena can be observed and appear as they would in a bulk system without too much additional unwanted effects due to small system size. If the objective would instead be to control the fluctuations around the average response, then the method of performing many simulations for each observation becomes compulsory; multiple responses are required to obtain such statistical information.

\subsection{The precipitate size distribution}

\begin{figure}[t!]
    \begin{center}
  \includegraphics[width=\columnwidth]{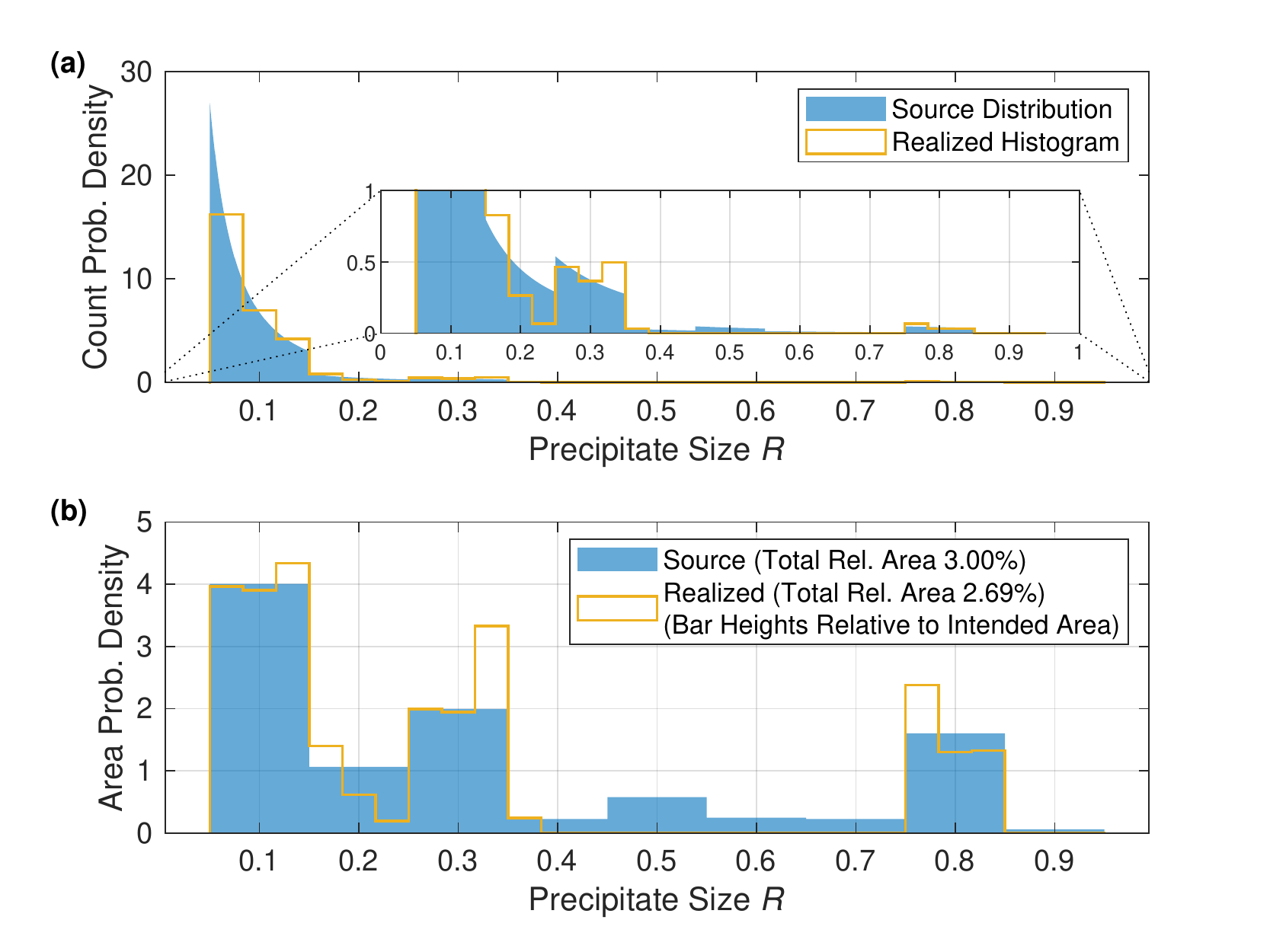}
  \caption{An example of a precipitate size distribution with respect to (a) the number of precipitates, (b) the area that the precipitates of each size cover. The staircase curve corresponds to a realized histogram made from an example set of size values drawn from the source distribution. This set corresponds to the sizes of the precipitates illustrated in Figure~\ref{fig:b1}. In this study, we attempt to optimize a nine-dimensional vector made of the relative piece heights (area portions) of the area-proportional, piecewise uniform source distribution. The objective is to maximize the average stress needed to cause a certain amount of strain.}
  \label{fig:b3}
\end{center}
\end{figure}

\noindent
Sizes for the precipitates are generated from a continuous, piecewise uniform size distribution that describes how the area of the simulation box that is reserved for precipitates is portioned among different precipitate sizes. There are nine adjacent pieces, each having a locally constant probability density. Each piece covers a size interval of width 0.1, and the centers of the pieces are evenly spaced between 0.1 and 0.9. The height values of the uniform pieces can be used to form a nine-dimensional vector. Our optimization problem is then to find the optimal vector with respect to a chosen objective derived from the stress-strain curve, leading to a designed response of the dislocation system.

As an example, Figure~\ref{fig:b3} shows the distribution from which the size values were drawn for the precipitates of Fig.~\ref{fig:b1}. As mentioned, our focus is on the area-proportional version of the distribution (Fig.~\ref{fig:b3}b) rather than the count-proportional (Fig.~\ref{fig:b3}a) because, for estimating the impact on the system behavior, the total mass of the precipitates is a better measure than their number. A constraint fixes the total precipitate area, so it is natural to have the size distribution describe how this area is divided among different precipitate sizes.

Precipitates are placed uniformly at random over the simulation box. Methods such as a minimal distance between precipitates could be used to mimic the effects of a manufacturing method on the precipitate arrangement. For simplicity, we do not impose any such additional requirements and do not prevent precipitates from overlapping. We do want to assume any particular manufacturing method because perhaps, in practice, the precise control of (the shape of) the precipitate size distribution requires some special manufacturing method instead of a common one. Also, the precipitate locations in the 2D simulation box could be viewed as projections of 3D locations where corresponding 3D spherical precipitates would not actually overlap (although the precipitates are assumed to be purely 2D for the precipitate-dislocation interaction model).

Let us attempt to make a quick estimate of what the precipitate sizes we use would correspond to in reality. The initial density of dislocations in our model is 0.04 per area unit (within the simplified unit system); this may slightly decrease during the simulation due to annihilations. Dislocation densities in metals vary in magnitude~\cite{ohring1995mechanical}, usually from $10^{12}$ to $10^{16}\;\mathrm{m/m^{3}}$. The size range of precipitates in this work starts from 0.05 and ends to 0.95. Using the dislocation densities as reference (choosing $10^{14}\;\mathrm{m/m^{3}}$ as the typical density), these sizes would in reality correspond to $1\;\mathrm{nm}$ and $19\;\mathrm{nm}$, which are quite typical sizes of real precipitates~\cite{fang2019statistical}.

\subsection{Generating precipitate sizes}

\noindent
The algorithm for generating precipitate sizes will be explained for the general $d$-dimensional case as the intention is to use it again for the 3D case in a future work. The algorithm is count-controlled; the number of generated precipitates is equal to the expected number (rounded to an integer) determined from the source distribution. The area which the precipitates occupy (3\% of the total area on average) may therefore slightly fluctuate between random configurations.

In detail, the precipitate size generator works as follows. The starting point is the piecewise uniform probability density with respect to area (or volume in three dimensional systems), represented by a vector consisting of piece heights. First, each vector element should be multiplied by the corresponding piece width (necessary only if the width varies) and then normalized so that the sum of the vector elements is 1. Now, each vector element corresponds to the area-proportional probability (mass) contained within each piece.

With respect to count, the probability density within a piece is not uniform (see Fig.~\ref{fig:b3}) but has density $\sim R^{-d}$, where $d=2$ is the spatial dimensionality of the dislocation model. To obtain a size $R$ from such distribution, we perform a conversion from area-proportional to count-proportional. The first step is to multiply (weight) each piece's (with edges $R_1$ and $R_2$; $R_1 < R_2$) probability mass by $\frac{\int_{R_{1}}^{R_{2}} r^{-d}\mathrm{d}r}{R_{2} - R_{1}}$ (which is the mean of the integrated function on the interval, because the integral gives the area, equal to the product of the mean and the interval of integration). If a piece's width is zero ($R_1 = R_2$), then the probability mass should be multiplied by $R_1^{-d}$ instead. Taking the sum over the results after the multiplication step gives the expected value $\langle R^{-d} \rangle$ which is useful for calculating the expected number of precipitates ($d=2$: $\langle N_{p} \rangle = \langle \frac{A_{p}}{\pi R^2} \rangle = \frac{A_{p}}{\pi} \langle R^{-2} \rangle$, where $A_{p}$ is the area reserved for precipitates). The count-proportional probabilities (probability masses) are then obtained by normalizing the previously obtained (weighted) results by their sum.

Now, one piece (a size interval) is selected randomly based on the count-proportional probabilities. Next, a size value is drawn within the selected interval from the distribution proportional to $R^{-d}$ (which gives uniform distribution with respect to covered area). The edges of the selected interval are first converted (mapped) into a representation where the distribution would be uniform. This can be done by operating with $f(x)=x^{-d+1}$ on the edges. Then, a number is drawn from the uniform distribution within the converted interval. Finally, the number is converted back by operating with $f^{-1}(x)=x^{\frac{1}{-d+1}}$ to obtain the size $R$.

\section{Method: Bayesian optimization}
\label{sec:bayes}

\noindent
The Bayesian optimization method~\cite{pelikan1999bayesian, mockus2012bayesian, frazier2018tutorial} utilizes the idea of Gaussian process regression where the objective function, from which the minimum (or maximum) is to be found, is represented as a probability distribution field over the search space. Each observation is assumed to be drawn from some Gaussian distribution; only the distribution parameters (mean and standard deviation, with respect to the objective value) are assumed to vary across the search space. The relation of how the parameters change when moving in the search space is modelled with a covariance function (a {\it kernel}). It tells how correlated, given a separation, the observations for different points of the search space are assumed to be. Utilizing this, the algorithm estimates the probability distributions of future observations based on past observations. The covariance function allows to inter- and extrapolate beyond observed points, providing estimates of future observations and their uncertainties at unexplored points of the search space. The estimated objective function obtained this way gives Gaussian distribution parameters (effectively the center and width) for every point of the search space. The width parameter in this case models the uncertainty due to noise (the only source of uncertainty for the exact underlying objective function) and also due to not having an observation at the exact location of the search space.

Based on previous observations, following inputs (for which the next observations will be obtained) can be chosen either by observing close to the current estimated optimum in hope of fast convergence or by focusing on the areas where the model still has large uncertainty, possibly discovering new potential locations for the global optimum. These two strategies are called exploitation and exploration, and an algorithm typically balances a mix between the two, starting by mostly exploring and then switching gradually towards exploitation. In practice, this process is controlled by an acquisition function which can be chosen from multiple options.

In this work, the MATLAB~\cite{matlab} implementation of the algorithm (the {\it bayesopt} function) is used, and the acquisition function is chosen to be {\it expected improvement}, a common choice found across many implementations. The covariance (kernel) function can be chosen freely too, but we use the default one in MATLAB called {\it ARD Matern 5/2}. This kernel uses {\it automatic relevance determination} which optimizes the length scale (a kernel variable) for each variable of the search space alongside the actual optimization process. Other options are also set to default values, except the {\it IsObjectiveDeterministic} option which is set to false. Also, a custom function is passed to the {\it ConditionalVariableFcn} option to enforce a norm equality constraint; further details are given in Section~\ref{sec:generation}.

The main advantage of Bayesian optimization over other optimization algorithms is that it can converge quickly relative to the number of observations. It is therefore useful for cases where observations are difficult to obtain, such as when each observation requires heavy computation or is expensive in some other way. Another advantage is that the algorithm works well for cases where the observations contain noise; there is usually a noise evaluation mechanism included, which takes the noise into account when estimating the objective function. Also, this method does not depend on whether the gradient of the objective function can be determined or not, and does not assume any particular shape for the feasible search space nor the objective function. One disadvantage is that the algorithm works well only for relatively low-dimensional input spaces, with an upper bound commonly set to 20 dimensions~\cite{frazier2018tutorial}. Although Bayesian optimization has a reputation of converging very quickly, having enough noise can inevitably raise the number of iterations needed for convergence.

\section{Generating feasible trial solutions}
\label{sec:generation}

\noindent
The aim is to optimize a continuous precipitate size distribution represented by a non-negative, nine-dimensional vector (see Section~\ref{sec:model}). The total area covered by the precipitates is fixed. This requirement can be formulated as a norm constraint, namely as a Lebesgue $p$-norm ($L^p$) equality constraint with $p=1$; the sum of the vector elements must be equal to a constant. Without loss of generality, the constant is here chosen to be 1. Next, we discuss different approaches for enforcing such constraints. To this end, the problem is first studied using mathematical notions before connecting the discussion back to the optimization problem. The approach starts by viewing trial solution vectors (let the number of elements be $n$) as points in an $n$-dimensional search space. (For our case with the nine-dimensional vectors, $n=9$.)

Bayesian optimization uses random sampling to explore the search space. In practice, this means that the algorithm generates sets of trial points uniformly at random over a hypercube for the purpose of selecting the most promising point according to the acquisition function, but every generated point may not be feasible as such due to constraints. The basic approaches for enforcing a constraint are either to reject the infeasible points or to map (modify) the points somehow to make them feasible. We will call these approaches the {\it rejection method} and the {\it mapping method}.

An important aspect is that feasible points should cover the feasible space evenly enough (uniformly in the optimal case). The reason is that if some regions of the space are over-represented and others under-represented, the optimization algorithm may be stuck to the over-represented areas and possibly unable to approach the global optimum in case it is located elsewhere.

For most problems, there is usually some rejection method that gives uniformly distributed feasible points. Rejection rate determines how many random trial points are needed on average to obtain one feasible point. A major drawback of this method is the {\it curse of dimensionality}; the probability of a random point being feasible tends to decay exponentially with respect to the number of dimensions $n$ in the case of norm constraints. Consequently, in high-dimensional problems, the required amount of random numbers that must be generated for the algorithm to work can become too high, forming a computational bottleneck. Therefore, this method is usually not suitable for other than low-dimensional problems.

In the case of a norm equality constraint, one simple mapping method is to divide every trial point (vector) by their norm (and then multiply by the required norm if it is something else than 1). This could be interpreted as radial projection onto the feasible hypersurface. Every mapped point would be feasible, but the mapping gives a non-uniform distribution of feasible points. Although this is not ideal, the method might still be useful if the symptoms are not severe. Later in Section~\ref{sec:results}, we demonstrate that this method causes the previously discussed scenario where the optimization algorithm cannot access certain regions of the search space (namely the boundaries and their proximity, consisting of sparse solutions) and therefore is unable to converge properly.

\subsection{From a hypercube to a hypersphere}

\noindent
In the following, we derive a mapping from the uniform distribution over the unit hypercube to another uniform distribution over the unit hypersphere. The mapping works generally for any number of dimensions $n$. Although at first glance it may seem useful only for problems having a Euclidean ($L^2$) norm equality constraint, we explain how to take advantage of the approach in the case of other $L^p$ norm constraints too.

Let $\mathbf{u} = \begin{bmatrix} u_1 \; u_2 \; \dots \; u_n \end{bmatrix}$ represent an $n$-dimensional vector and $||\cdot||_p$ be the notation for the $L^p$ norm, which for $\mathbf{u}$ corresponds to $||\mathbf{u}||_p = (\sum_{i=1}^n |u_i|^p)^{1/p}$. The feasible space in our case is the space formed by all such $\mathbf{u}$ for which $||\mathbf{u}||_1 = 1$. For example, in three dimensions, the shape of the space is a regular octahedron. The non-negativity requirement restricts the search space to one of its (hyper)faces, but even so, the shape is quite nontrivial in high dimensional spaces~\cite{ahmadi2019uniform}. In contrast, the $L^2$ norm equality constraint $||\mathbf{u}||_2 = 1$ corresponds to the unit hypersphere (the surface of the unit hyperball) which is mathematically easier to deal with.

It is possible to first find a mapping for the $L^2$ problem and then use it for the $L^1$ problem. The trick is to change the problem definition so that the objective is not to find the probability distribution itself but a {\it probability amplitude function}, something that resembles the wave functions in the context of quantum physics (although non-negative real-valued functions suffice here). The $L^1$-constrained probability distribution can be obtained from the $L^2$-constrained probability amplitude function simply by squaring the magnitude of each amplitude value. If this operation is viewed to belong to the objective function, the original $L^1$-constrained problem effectively changes into an $L^2$-constrained problem. A similar approach could be applied generally to other $L^p$ constraints by raising the magnitudes to the power of $2/p$ instead of $2$. (Formally, the hypercube itself corresponds to an $L^\infty$ constraint when symmetrically centered at the origin.)

It is well known~\cite{muller1959note} that the multidimensional standard normal distribution is isotropic. A sample from this distribution can be derived by drawing $n$ independent samples from the standard normal distribution and by forming a vector of them. The $L^2$ norm equality constraint can be satisfied by dividing such vectors by their $L^2$ norm. Isotropy is preserved, so the distribution of these resulting vectors over the hypersphere is uniform.

The final missing link is to find a way to convert random numbers drawn from the standard uniform distribution into numbers that follow the standard normal distribution. This can be done by operating with the \emph{quantile function} (the inverse function of the cumulative distribution function) of the standard normal distribution.

Let $\mathbf{x}$ be an $n$-dimensional vector representing a location inside the unit hypercube (vector elements can take values between 0 and 1). Now, the previously discussed steps can be combined taken to obtain a corresponding point $\mathbf{u}$ on the unit hypersphere. The result is a mapping from the hypercube onto the hypersphere which preserves uniformity:
\begin{equation}
\label{eq:mapping}
\mathbf{u}(\mathbf{x}) = \frac{\Phi^{-1}(\mathbf{x})} {||\Phi^{-1}(\mathbf{x})||_{2}},
\end{equation}
where $\Phi^{-1}$ is the quantile function of the standard normal distribution (applied element-wise).

Additionally, non-negativity for the elements of $\mathbf{u}$ can be enforced by scaling the hypercube to the interval $[0.5,1)$ in each dimension or by taking the absolute values of the resulting points. Due to the properties of the quantile function, element values of $\mathbf{x}$ should not be exactly 0 or 1 (or if necessary, such cases should be considered separately). The mapping is from a higher dimensional space to an effectively lower dimensional hypersurface, and information worth of one vector component is lost, so there is no inverse function.

There are many existing alternative methods for generating points uniformly at random over a hypersphere~\cite{muller1959note, marsaglia1972choosing, guralnik1985algorithm, harman2010decompositional}. Our approach takes advantage of the properties of the normal distribution; one alternative suggestion~\cite{guralnik1985algorithm} generalizes the idea of the polar coordinate system. We have now shown that the basic idea of these generators can be extended to develop mappings, hopefully enabling new problem solving capabilities.

\subsection{From a hypercube to a hyperball and vice versa}

\noindent
It is also possible to map the hypercube (the $n$-cube) uniformly and continuously into the hyperball (the $n$-ball). Let $\mathbf{v}$ be an $n$-dimensional vector. The unit $n$-ball corresponds to the inequality constraint $||\mathbf{v}||_2 < 1$. In this case, the dimensionality of the space is not changed, so a corresponding one-to-one inverse mapping also exists. Although our current problem only requires dealing with an equality constraint, many other problems involve an inequality constraint, so it is good to consider both of these related cases here.

The mapping for the case of an inequality constraint can be derived by using these facts:
\begin{itemize}
    \item A distribution with some probability density function $f$ can be mapped into the standard uniform distribution by operating with the cumulative distribution function $F$ corresponding to $f$. Correspondingly, the quantile function $F^{-1}$ maps the standard uniform distribution into the distribution $f$.
    \item Vector $\mathbf{p}$ formed of $n$ independent standard normal distributed variables follows an isotropic~\cite{muller1959note} multivariate distribution, and the sum of squares ($||\mathbf{p}||_2^2$) follows the $\chi^2$-distribution.
    \item The radial coordinate $q_r = ||\mathbf{q}||_2$ of a point $\mathbf{q}$ drawn from a spherical uniform distribution in $n$ dimensions follows a distribution $f(q_r) \sim q_r^{n-1}$, and therefore, for the uniform distribution over the unit $n$-ball, $F(q_r)=q_r^n$ with $q_r \in [0,1)$.
\end{itemize}
Now, putting everything together gives for each vector $\mathbf{x}$ of the unit $n$-cube the corresponding vector $\mathbf{v}$ within the unit $n$-ball:
\begin{equation}
\label{eq:mappingC2B}
\mathbf{v}(\mathbf{x}) = \frac{\Phi^{-1}(\mathbf{x})} {||\Phi^{-1}(\mathbf{x})||_{2}} [F_{\chi^2}(||\Phi^{-1}(\mathbf{x})||_{2}^{2};\,n)]^{1/n},
\end{equation}
where $\Phi^{-1}$ is the quantile function of the standard normal distribution (applied element-wise), $F_{\chi^2}$ the cumulative distribution function of the $\chi^2$-distribution and $n$ the number of dimensions (the number of vector elements). Alternatively, without changing the outcome, one could replace the $\chi^2$-distribution with the $\chi$-distribution along with giving the norm instead of the norm's square as input for the cumulative distribution function.

\begin{figure}[t!]
    \begin{center}
  \includegraphics[width=\columnwidth]{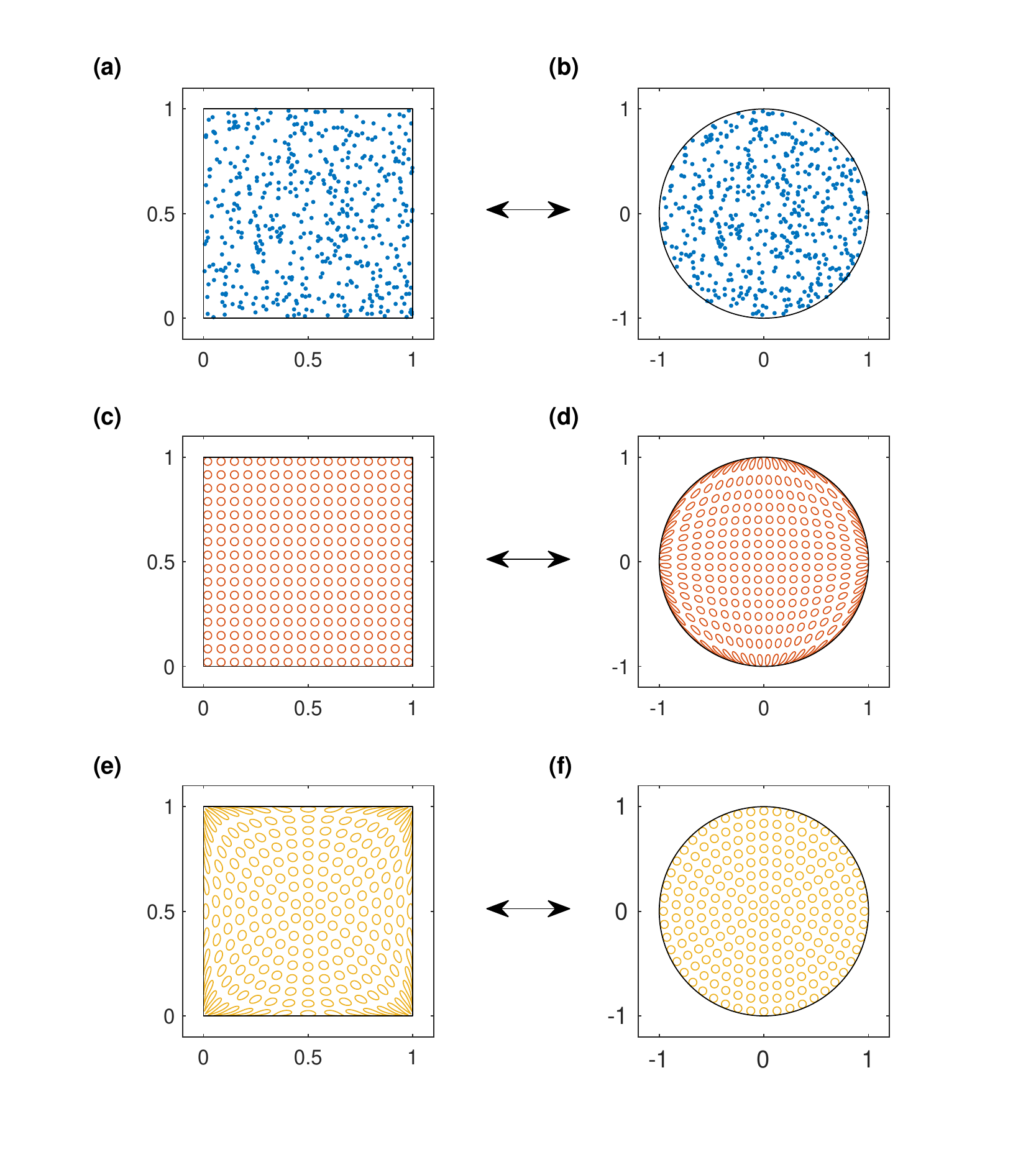}
  \caption{Two-dimensional mapping examples between a square (2-cube) and a disk (2-ball). Eq.~(\ref{eq:mappingC2B}) maps the left-side to the right-side, and Eq.~(\ref{eq:mappingB2C}) does the reverse. The first row (\textit{a} and \textit{b}) shows uniformly distributed random points; the regular pattern of \textit{c} maps into \textit{d}, and \textit{e} corresponds to the regular pattern of \textit{f}. The uniformity and continuity of the mapping means that the ratio of hypervolumes (areas in 2D) of any two subsets remains the same after the mapping.}
  \label{fig:mapping}
\end{center}
\end{figure}

As the function of Eq.~(\ref{eq:mappingC2B}) is injective (one-to-one), the operation can be reversed:
\begin{equation}
\label{eq:mappingB2C}
\mathbf{x}(\mathbf{v}) = \Phi\left(\frac{\mathbf{v}} {||\mathbf{v}||_{2}} [F_{\chi^2}^{-1}(||\mathbf{v}||_{2}^{n};\,n)]^{1/2}\right),
\end{equation}
where the cumulative distribution function $\Phi$ of the standard normal distribution is applied element-wise, and $F_{\chi^2}^{-1}$ is the quantile function of the $\chi^2$-distribution. Examples of applying the mappings for $n=2$ are demonstrated in Figure~\ref{fig:mapping}.

Although many alternatives exist~\cite{guralnik1985algorithm, harman2010decompositional}, there does not seem to be a widely known method to generate points uniformly at random over a hyperball with Eq.~(\ref{eq:mappingC2B}) that would correspond to our proposed way. One common suggestion~\cite{harman2010decompositional} starts similarly to the hypersphere case, but the final radial coordinate is obtained from an additional random number that is raised to the power of $1/n$, so the approach requires $n+1$ random numbers to be generated. In comparison, our method utilizes the distribution conversion technique (the first item in the previous bulleted list), which seems to be quite uncommon in existing literature. As a result, Eq.~(\ref{eq:mappingC2B}) takes only $n$ numbers, and it can be interpreted not just as a generator but as an injective mapping from a hypercube to a hypersphere, paired with the inverse mapping of Eq.~(\ref{eq:mappingB2C}).

\subsection{Solving constrained optimization problems with the mappings}

\noindent
How to enforce the norm equality constraint on high-dimensional trial solutions so that there are no convergence problems or computational bottlenecks? Finding a solution to this problem was the original motivation for deriving these mappings. There are many different types of constraints, each requiring a specific method for handling them. Instead of providing a method for every possible constraint, implementations of optimization algorithms usually rely on the end user to come up with the method, which is then used in tandem with the rest of the algorithm.

A recent implementation of the Bayesian optimization algorithm in MATLAB~\cite{matlab}, which was used for this study, generates trial solutions by forming vectors from independent random numbers, each following a uniform distribution. Such vectors (here corresponding to $\mathbf{x}$) do not satisfy the norm constraint as such, but there are ways to enforce it. We do this by writing (programming) a function that is given to the {\it bayesopt} command through the {\it ConditionalVariableFcn} option. This function allows to modify (or to perform a feasibility check on) the trial vectors after they are created but before they are used as actual candidate solutions in the optimization process.

As explained previously, a simple way would be to divide each vector $\mathbf{x}$ by their norm, but this would cause convergence problems (which will be demonstrated in Section~\ref{sec:results}). Instead, our conditional variable function implements Eq.~(\ref{eq:mapping}), enforcing the $L^2$ norm equality constraint by mapping vectors $\mathbf{x}$ into vectors $\mathbf{u}$ (with $n=9$). The elements of $\mathbf{x}$ have values within $[0.5,1)$ to assure that the elements of $\mathbf{u}$ are non-negative. (If we were to deal with a norm {\it inequality} constraint instead, the conditional variable function would implement Eq.~(\ref{eq:mappingC2B}) in the place of Eq.~(\ref{eq:mapping}), and the result of the mapping would be $\mathbf{v}$ in this case.) Also, the beginning of the objective function is modified so that it raises the vector elements of $\bf u$ to the power of 2 (giving $\mathbf{u}^{\circ 2}$ in Hadamard notation). This {\it raising operation} is done in order to obtain the actual precipitate size distribution that follows the required $L^1$ norm equality constraint before proceeding to the DDD simulations (the main task of the objective function). (If we would be dealing with other than just non-negative values, the raising operation should only raise the magnitudes (absolute values) to the desired power and not change the signs of the vector elements).

The mapping and the raising operation could, however, be implemented into different stages of the algorithm, but these choices are not equivalent. The optimization algorithm sees (and gives the estimated solution in) the representation that is given by the conditional variable function and passed on to the objective function. (The $L^1$-constrained representation $\mathbf{u}^{\circ 2}$ will always be used when we visualize a solution.)
\begin{itemize}
    \item If both the mapping and the raising operation are performed in the conditional variable function, (Euclidean) distances between trial solutions (for the purposes of the Bayesian optimization algorithm) are measured in the $L^1$-constrained representation $\mathbf{u}^{\circ 2}$ (or $\mathbf{v}^{\circ 2}$).
    \item If the mapping is performed in the conditional variable function and the raising operation within the objective function (as we choose to do here), distances are measured in the $L^2$-constrained representation $\mathbf{u}$ on the hypersphere (or $\mathbf{v}$ in the hyperball).
    \item If both the mapping and the raising operation are performed within the objective function, distances are measured in the unprocessed representation $\mathbf{x}$ in the hypercube.
\end{itemize}
Different representations belong to different search spaces. Consequently, changing the effective representation affects the distances between the trial solutions, which is a relevant aspect for the kernel's functioning (see Section~\ref{sec:bayes}).

The third choice should be avoided in the case of an equality constraint as the hypercube is dimensionally larger than necessary. However, it could be utilized in problems having an inequality constraint to effectively get rid of the constraint. As mentioned, we use the second way to obtain the results presented in Section~\ref{sec:results} (the only exception being Figure~\ref{fig:b4} where an alternative mapping method is tested). This choice is based on the idea that the vectors $\mathbf{u}$ are uniformly distributed over the hypersphere. If instead the vectors $\mathbf{u}^{\circ 2}$ would be distributed uniformly (both $\mathbf{u}$ and $\mathbf{u}^{\circ 2}$ cannot be at the same time), the first choice could be the better one (intuitively). Perhaps the shape of the search space also affects convergence in the sense that round, isotropic ($L^2$ related) search spaces are possibly the most suitable spaces because they do not have corners or anything that would give a bias towards some vector direction over others.

\section{Results}
\label{sec:results}

\noindent
As a simple test case, we attempt to find the precipitate size distribution giving the maximum expected stress at strain $\epsilon=0.2$, starting with the case of the precipitate potential of Eq.~(\ref{eq:R1potential}). The expected stress of a finite system works as an estimate of the response of a bulk system. If assuming that the optimum distribution is a $\delta$ distribution, its location is known to be at the boundary of the search space (Fig.~\ref{fig:b2}a). This information helps when testing how different approaches and parameter choices affect the convergence of the optimization. To obtain general convergence results, we do not place the initial points (trial solutions) at or near the boundaries of the search space (which would be the suggested way), but instead start from a set of randomly placed points.

The optimization objective is traditionally minimized; any maximization problem can be turned into a minimization problem by inverting the objective's sign. Following this common practice, the objective is here defined to be the negative of the stress (at strain $\epsilon=0.2$). This way, curves illustrating convergence (of the estimated objective) are decreasing, which is consistent with the majority of other optimization problems across publications.

\begin{figure}[t!]
    \begin{center}
  \includegraphics[width=\columnwidth]{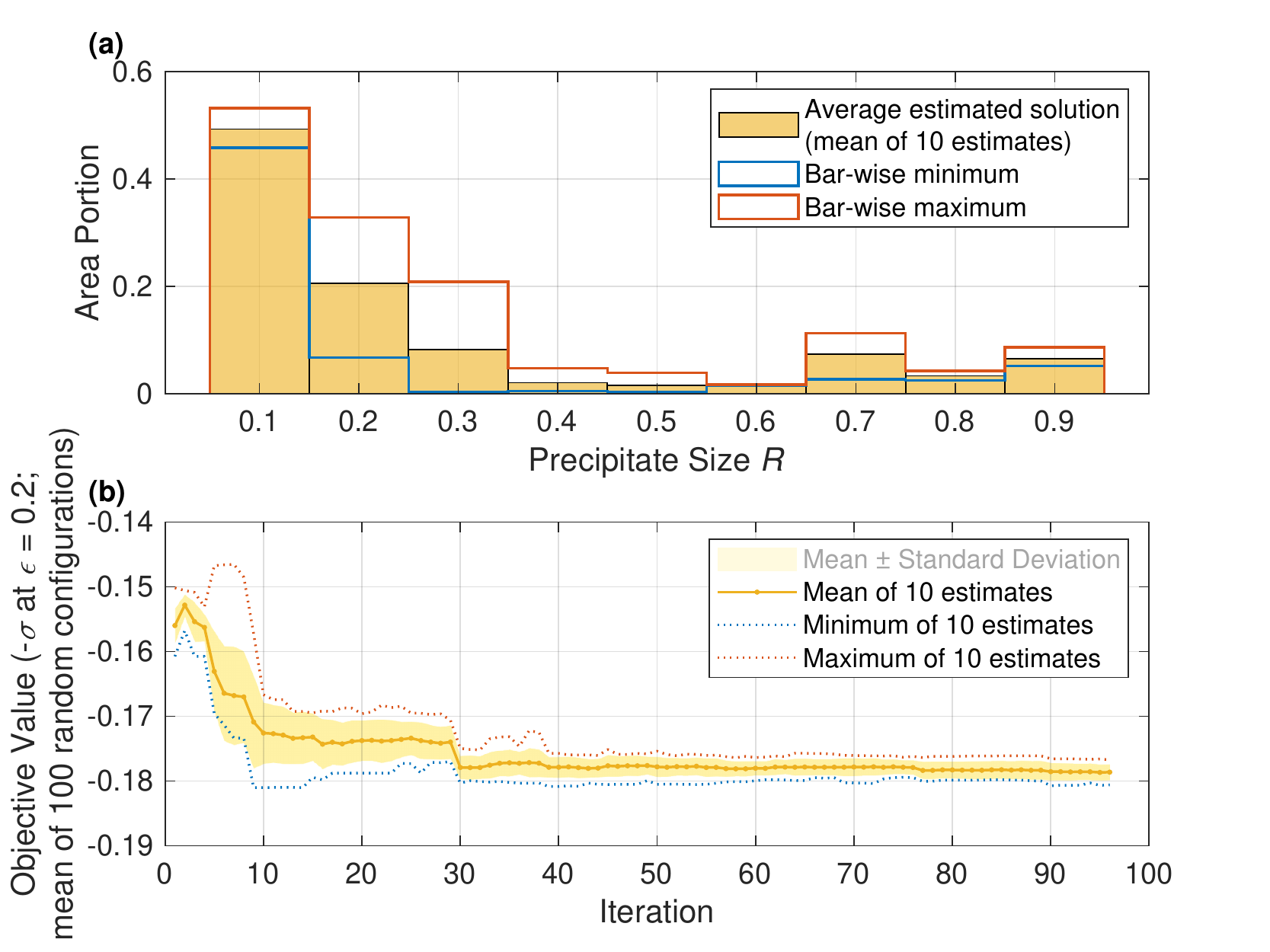}
  \caption{Bayesian optimization solution (a) and objective evolution (b) for the potential of Eq.~(\ref{eq:R1potential}). Each observation is an average over $M=100$ random configurations, and the search space is sampled simply by dividing random vectors from the unit hypercube by their $L^1$ norm. The optimization process was repeated 10 times with different random number seeds to obtain the statistical measures.}
  \label{fig:b4}
\end{center}
\end{figure}

\begin{figure}[t!]
    \begin{center}
  \includegraphics[width=\columnwidth]{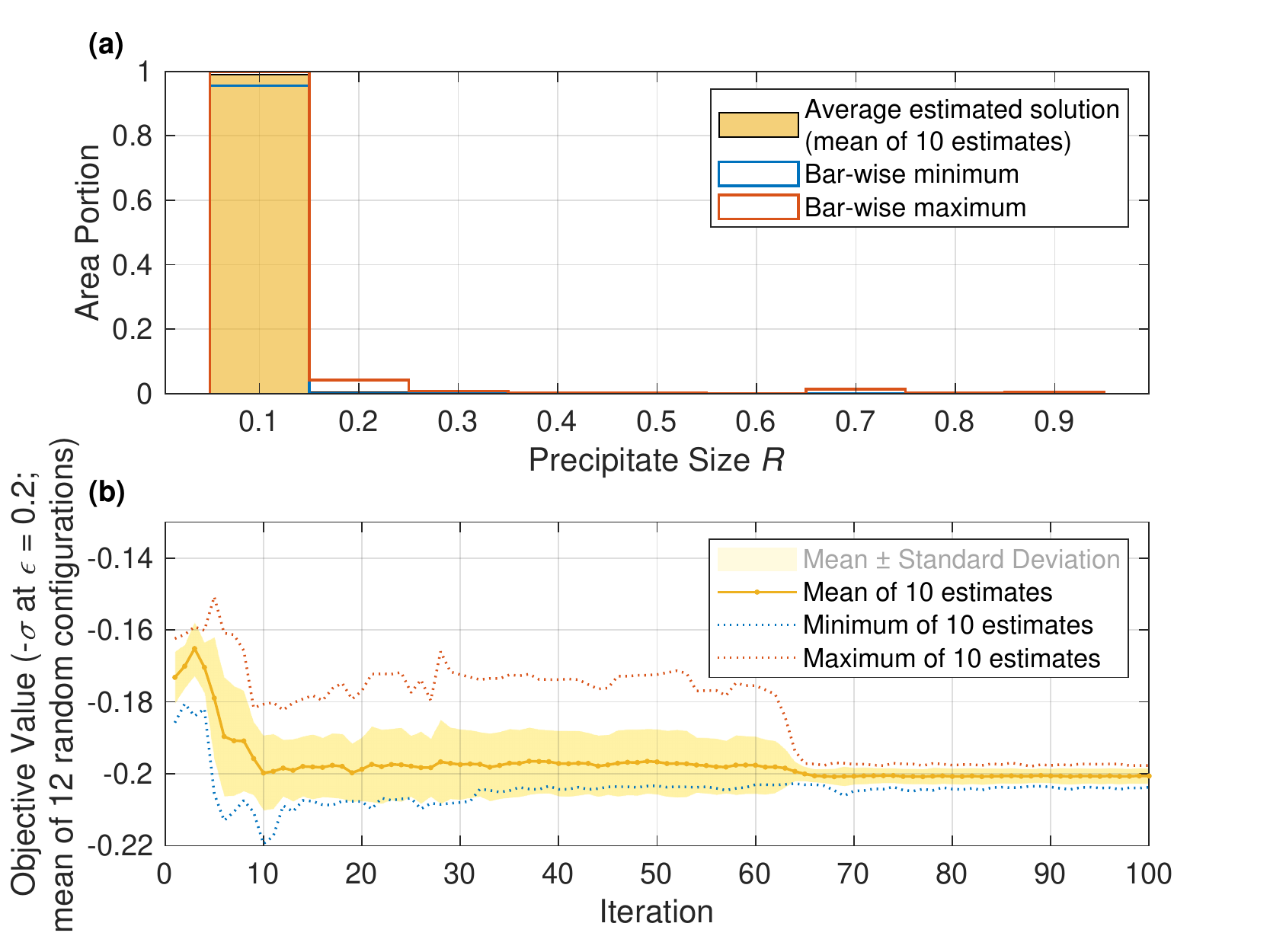}
  \caption{Bayesian optimization solution (a) and objective evolution (b) for the potential of Eq.~(\ref{eq:R1potential}). Each observation is an average over $M=12$ random configurations, and the search space is sampled by mapping random vectors from the unit hypercube onto the unit hypersphere with Eq.~(\ref{eq:mapping}), followed by squaring the vector elements within the objective function. The optimization process was repeated 10 times with different random number seeds to obtain the statistical measures.}
  \label{fig:b5}
\end{center}
\end{figure}

\begin{figure}[t!]
    \begin{center}
  \includegraphics[width=\columnwidth]{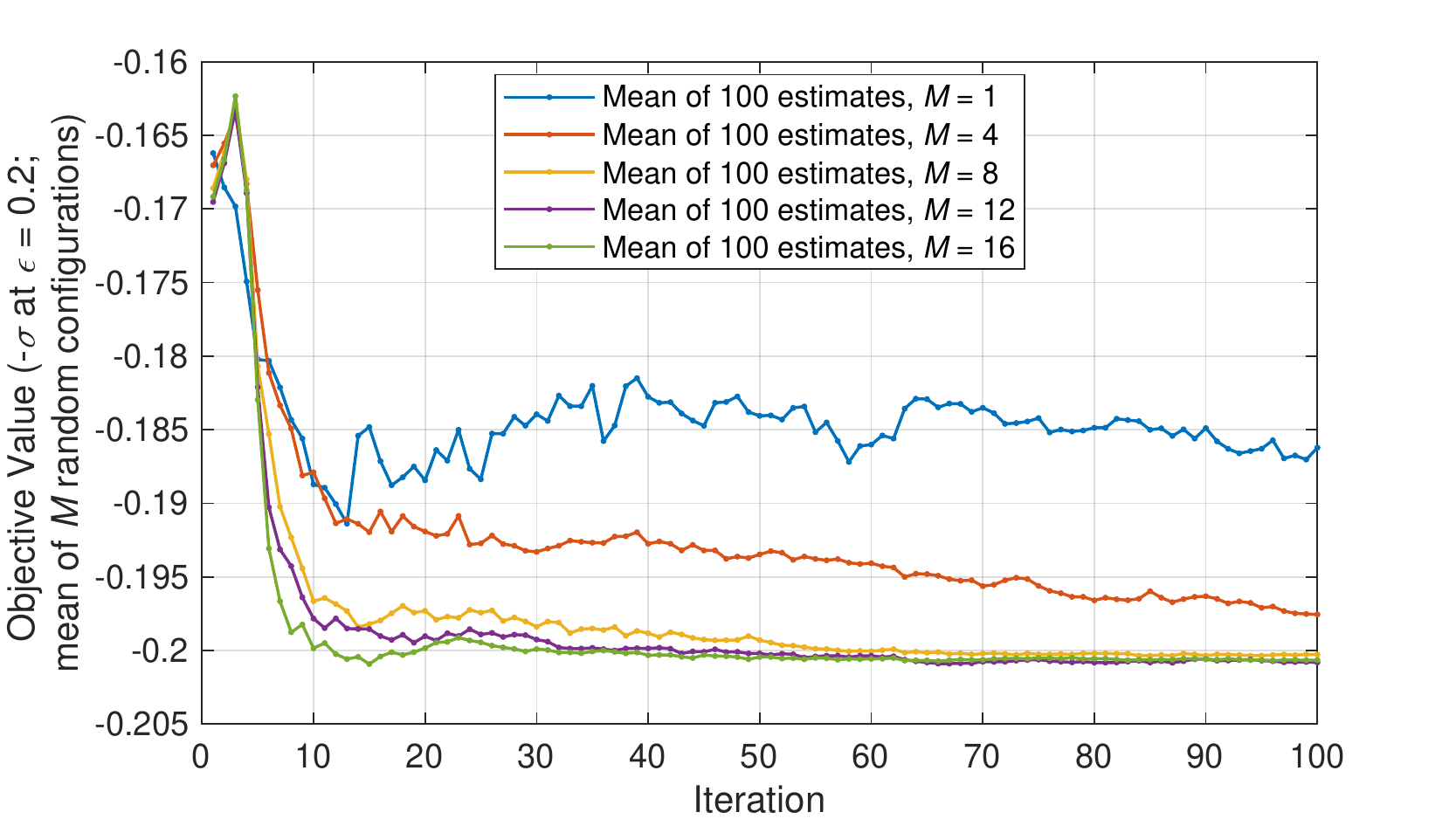}
  \caption{Evolution of the Bayesian optimization objective (negative of the stress, $-\sigma$, at strain $\epsilon=0.2$) for 100 optimization iterations for five different $M$. $M$ determines the number of simulations calculated for each trial solution with different random configurations. The average result over these simulations is used as the optimization objective in order to reduce noise. Here, the potential of Eq.~(\ref{eq:R1potential}) and the mapping of Eq.~(\ref{eq:mapping}) were applied. Optimization processes were repeated 100 times with different random number seeds, and shown here are the average curves.}
  \label{fig:b6}
\end{center}
\end{figure}

\begin{figure}[t!]
    \begin{center}
  \includegraphics[width=\columnwidth]{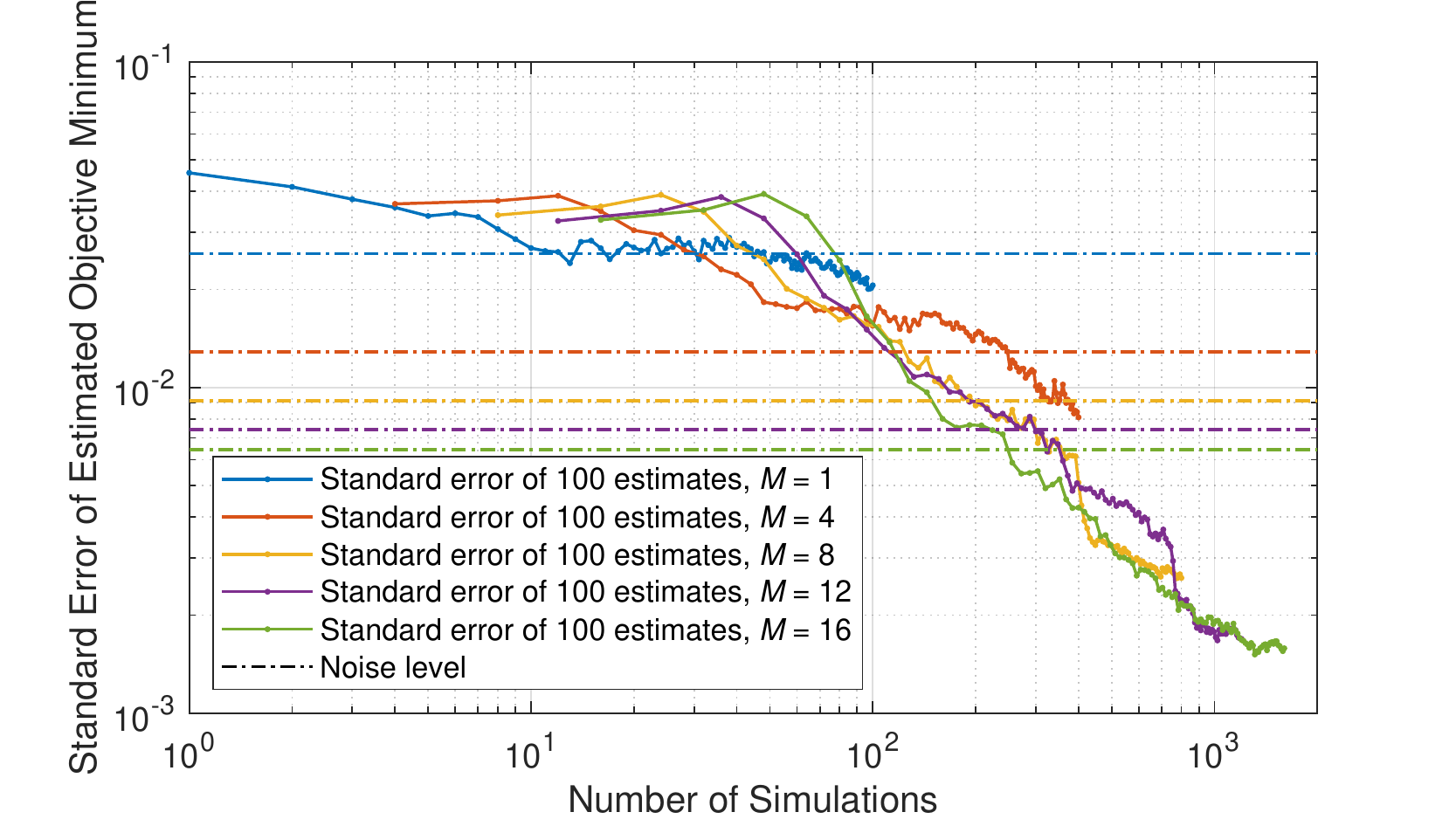}
  \caption{Convergence of the Bayesian optimization, measured as the standard error of the objective value of the estimated solution from that of the optimal solution, as a function of the number of simulations ($M$ times the iteration index), along with estimates of the noise level for each $M$. Here, the potential of Eq.~(\ref{eq:R1potential}) and the mapping of Eq.~(\ref{eq:mapping}) were applied. Optimization processes were repeated 100 times with different random number seeds, and the standard errors were calculated for the distributions that formed as a result.}
  \label{fig:b7}
\end{center}
\end{figure}

\begin{figure}[t!]
    \begin{center}
  \includegraphics[width=\columnwidth]{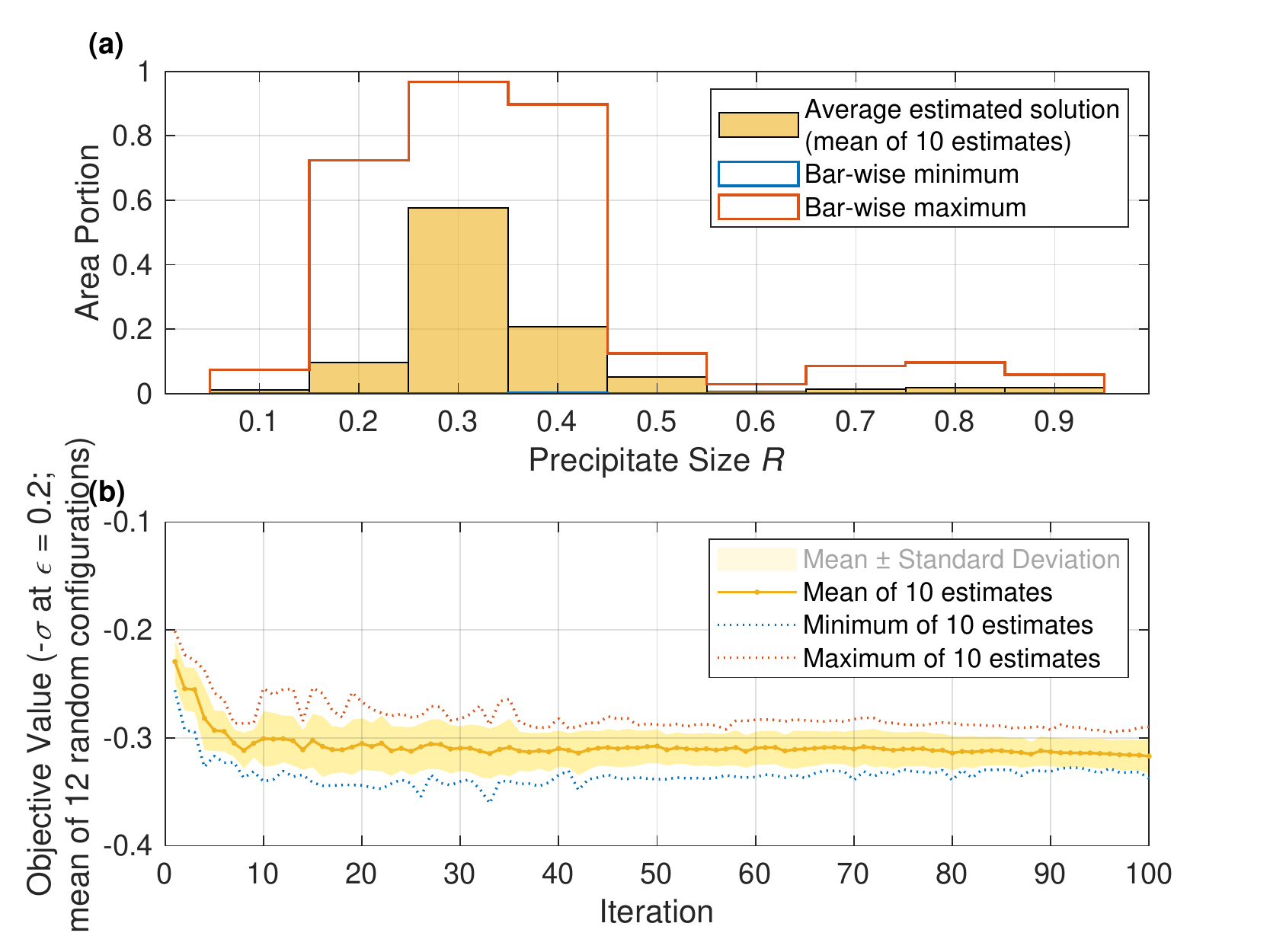}
  \caption{Bayesian optimization solution (a) and objective evolution (b) for the alternative potential of Eq.~(\ref{eq:R2potential}). Each observation is an average over $M=12$ random configurations, and the search space is sampled with Eq.~(\ref{eq:mapping}). The optimization process was repeated 10 times with different random number seeds to obtain the statistical measures.}
  \label{fig:b8}
\end{center}
\end{figure}

\begin{figure}[t!]
    \begin{center}
  \includegraphics[width=\columnwidth]{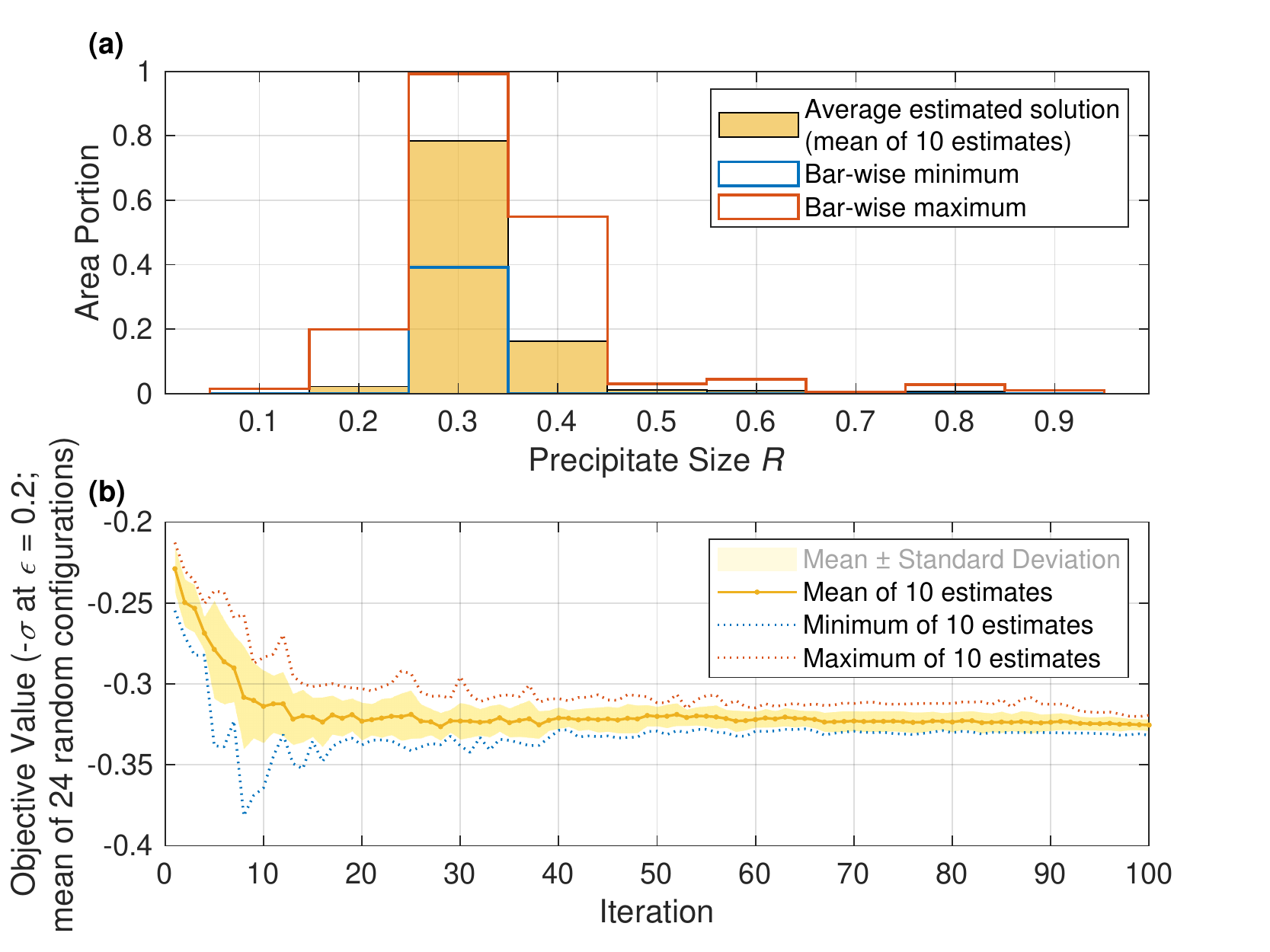}
  \caption{Bayesian optimization solution (a) and objective evolution (b) for the alternative potential of Eq.~(\ref{eq:R2potential}). Each observation is an average over $M=24$ random configurations, and the search space is sampled with Eq.~(\ref{eq:mapping}). The optimization process was repeated 10 times with different random number seeds to obtain the statistical measures.}
  \label{fig:b9}
\end{center}
\end{figure}

Positions of precipitates and dislocations are random, and realized sets of size values drawn from a precipitate size distribution can also vary slightly. These cause considerable deviations (noise) in the observed stress values between different random configurations. As mentioned previously in this work, Bayesian optimization works even with noisy objectives, but noise still slows down the convergence of the optimization. If there is too much noise, features of the underlying objective function, including the global optimum, may be hidden under the noise, preventing convergence. As a simple remedy, simulations are performed for multiple ($M$) random configurations, and the optimization algorithm sees the sample mean of those stresses as the observation for each candidate solution (that corresponds to a certain precipitate size distribution).

Initially, when using a simple method to enforce the constraint that fixes the total precipitate area, the solution did not converge towards a single piece distribution. It could be determined from the objective value (by comparing with Fig.~\ref{fig:b2}) that the resulting distribution was not as good as the single piece solution would be. The cause was unknown, so various optimization options, such as the acquisition function, were adjusted to see if they could make a difference. As the result did not change, we tried to increase $M$ to reduce noise considerably. Figure~\ref{fig:b4} shows the optimization result for $M=100$; there was no notable difference to the results for smaller choices of $M$. Finally, we tried to rethink alternative ways to enforce the constraint and came up with the method described in Section~\ref{sec:generation}). With the proposed method, the solution converges to a single piece with a significantly better objective value than before, even for a relatively small $M$ as seen in Figure~\ref{fig:b5}.

Figures~\ref{fig:b4}~and~\ref{fig:b5} show how important it is to choose a proper way for sampling the search space. When using a simple division by the $L^1$-norm to obtain feasible trial solutions from unconstrained vectors, the optimization algorithm is unable to reach the corners and boundaries of the search space due to uneven sampling density. As mentioned, reducing noise by using a larger $M$ does not resolve this issue. These problems can be avoided only by switching to a proper constraint enforcing method like those presented in Section~\ref{sec:generation}.

Figures~\ref{fig:b6}~and~\ref{fig:b7} illustrate how the convergence depends on $M$. It seems that if $M$ is not high enough, noise slows down or even prevents convergence, and there remains a gap between estimated and true objective minimum values. In such cases, the gap's size is about the same order as the noise level. For this specific optimization problem, $M\ge8$ seems to be sufficient for efficient convergence that can break through the limit set by the noise level.

According to Figure~\ref{fig:b7}, convergence becomes faster when using a larger $M$, although the computational cost of a single iteration increases in return. These balance out each other, so reaching a given proximity around the optimum has about the same total computational cost independent of $M$.

We also tested optimizing otherwise the same problem but now with the alternative precipitate potential defined in Eq.~(\ref{eq:R2potential}), and the results are shown in Figures~\ref{fig:b8}~and~\ref{fig:b9}, the first for $M=12$ and the latter for $M=24$. It could be that the optimal solution is again a single piece as the solution seems to converge towards a narrower distribution with respect to larger $M$. Unlike in the previous case with Eq.~(\ref{eq:R1potential}) as the potential choice, this time higher $M$ still continues to improve the solution, probably since the optimum is less prominent (see Fig.~\ref{fig:b2}b), meaning the objective value is almost the same no matter if the size distribution has some width or not.

Increasing $M$ means increasing the computational cost of the optimization, but the cost should preferably be minimized. A good choice for $M$ would be as small as possible that still gives a solution that has converged enough; it could be problem-specific. Having to guess a perfect value for $M$ could be avoided if it is possible to determine the limit solution from a series of low-$M$ solutions, like it seems to be with Figures~\ref{fig:b8}~and~\ref{fig:b9}. Starting from a small $M$ and then gradually increasing it (depending on the necessity) could be an efficient strategy for solving these kinds of problems.

\section{Discussion}
\label{sec:discussion}

\noindent
Optimal size distributions for our systems seemed to be narrow, possibly $\delta$ distributions, which could be considered as trivial solutions. Such solutions could be expected when searching for a linear combination of solution parameters which each have an independent effect on the objective value. In such cases, optimal (probability) weighting concentrates at the parameter having the best individual performance. This could be assumed true for the 2D system with point dislocations that have only one precipitate passing mechanism. It is not yet certain if this applies also to the 3D case with curved line dislocations and more mechanisms of getting past precipitates. Resolving this is intended to be one of the subjects of our following research.

We also considered minimizing the fluctuations of the stress-strain curves (either by minimizing the standard deviation of stress, or by reducing the size of avalanches), but found out that, in contrast to the average stress value, the standard deviation remains nearly constant with respect to precipitate size when testing with $\delta$ size distributions and using the potential of Eq.~(\ref{eq:R1potential}), so there would not have been much to optimize. The situation may change when moving to 3D models, so fluctuation minimization could then become a objective.

The effects of the precipitates were modelled by a Gaussian potential, assuming either linear (Eq.~(\ref{eq:R1potential})) or quadratic (Eq.~(\ref{eq:R2potential})) scaling for the interaction strength as a function of the radius. Another modelling option would be to use impenetrable precipitates~\cite{liu2021dislocation}, effectively assuming high interaction strength regardless of the precipitate size. There has been some comparison between molecular dynamics (MD) and DDD simulations with Gaussian interaction~\cite{lehtinen2016multiscale} but no extensive studies about which functional form for the interaction would best match with the MD results. Finding this out could be a possible direction for future research. Although many hardening effects are similar regardless of the functional form, optimization results may depend on it, especially on the scaling of the interaction strength with respect to precipitate size.

\subsection{Possible reformulations of the optimization problem}

\noindent
To increase the amount of control even more, precipitates could be allowed to have different shapes. Anisotropic shapes like ellipsoids would also make orientation a relevant property. Additional parameters could be defined for determining distributions over different shapes and orientations. If a new problem definition for our system involves a new set of constraints, we are likely to have the necessary tools ready thanks to the mappings derived in Section~\ref{sec:generation}. Both equality and inequality constraints for the norm can be satisfied without convergence problems by choosing a suitable mapping for generating feasible input points.

Optimization traditionally gives only the optimum choice of parameters and the corresponding objective value as the main result. One interesting question would be to ask how quickly the objective value deviates from the optimal value if the optimal parameter values were to be adjusted. This could be defined in many ways. One way would be to apply a constraint that sets a region around (and at) the optimum infeasible and solve the new problem, then repeating with gradually increasing the size of the infeasible region. Another way is for the case where each choice of input parameters is associated with a cost: a new constraint would be applied like in the previous case, but the constraint now sets an upper limit for the cost instead of a lower limit for distance from the original optimum. New optima would be searched for cost choices corresponding to fractions of the cost of the unconstrained optimum. Both proposed methods would result in decreasing curves showing the optimal objective value as a function of increasing displacement or decreasing cost.

Relying on Bayesian optimization limits the dimensionality of the search space~\cite{frazier2018tutorial}. As a remedy for the limitations this causes, one could perform a series of optimizations. If the optimal distribution appears to be narrow compared to the resolution of the discretized distribution, the domain of the distribution could be adjusted, now focusing on a smaller range and consequently having better resolution. Also, alternatives to the proposed size distribution format could be used, such as a similar one but with adjustable piece edges.

In true materials, precipitate size distributions are often controlled by processing the material in a way that leads to Ostwald ripening~\cite{mohles1999simulation, fang2019statistical} (condensation of small precipitates into larger ones), which involves control parameters. One idea is that these parameters could form the search space instead of the actual resulting size distribution, although the number of parameters could be very low (only time and temperature), and there may not be as much freedom to control the outcome. In that case, precipitate sizes tend to form a log-normal distribution as a result of condensation~\cite{fang2019statistical}. The process also results in a non-uniform spatial distribution of the precipitates, which could have some effect on the material response compared to a random configuration. Still, directly optimizing the size distribution perhaps generalizes better for different applications, because it does not assume any specific processing method.

\subsection{Other uses for the mappings}

\noindent
The mappings proposed in Section~\ref{sec:generation} give a way to convert between spaces (or domains) having different geometrical shapes while preserving many important properties, and they work generally for any number of dimensions. Here, the mappings were proposed as workarounds for enforcing constraints, but they could be useful for many other potential uses as well.

Instead of gradually adjusting explored points based on observations, generating sets of random trial points to sample the search space and selecting the best candidate based on the value of the acquisition function is common in Bayesian optimization algorithms. This is a good method (given that the density is distributed well over the feasible space) especially when exploring the space. But, when exploiting, this method depends on the assumption that with a high enough probability, among the set of generated trial points, there is a point close enough to the optimum so that the next iteration would improve the solution. With the proposed mappings, it should be possible to control the input directly based on the feedback. Normally, it might be difficult to move in a feasible space that has an exotic geometrical shape, having to constantly worry about feasibility. In contrast, hypercubes have a simple shape, and feasibility of a vector can be determined element-wise without having to check the values of other elements. By using a mapping to redefine the problem, one could easily adjust the hypercube representation of an estimated solution towards the most promising direction based on the acquisition function. This idea could potentially turn out to be a good way to boost optimization efficiency for some problems. However, the advantage may be limited if noise is the main bottleneck limiting optimization convergence.

Thinking about other applications, the mappings could serve as tools for conversion between different representations of data. One could map a coordinate system of one representation to another, giving as a result a curved coordinate system with new properties. It is also possible to draw random samples from (or create mappings for) other uniform distributions with related geometrical shapes like cylinders.

\subsection{Conclusion}

\noindent
Bayesian optimization was utilized to find the optimal precipitate size distribution that maximizes the expected stress needed to move dislocations by a given strain value in 2D DDD models. The study revealed technical challenges associated with constraints and noise. A constraint could be applied in many ways, but the proposed mapping method was found to fit well with the optimization algorithm, considerably improving convergence and removing computational bottlenecks when comparing with other methods. Averaging over many random configurations can reduce the noise without increasing computation time when the simulations are done in parallel.

Optimization results were evaluated and justified by comparing with statistics from systems with monosized precipitates. Also, by repeating optimization processes for different random number seeds, we were able to reveal the statistical nature of the optimization convergence. Such thorough results can be calculated quite effortlessly when dealing with simplified, low-dimensional DDD models. Results suggest that optimal precipitate size distributions tend to resemble $\delta$ distributions even when no particular shape is assumed for the distribution. More importantly, these findings provide preparations for moving to larger scale, more realistic 3D models where, due to a larger variation of bypassing mechanisms, results may not be as trivial. Optimization of such systems is expected to bring similar technical challenges related to constraints and noise. Not only sizes, but (the distributions over) precipitate shapes and orientations could also be optimized. The design objective could also be chosen differently. One interesting direction would be to minimize strain bursts, the main cause of fluctuations between the stress responses of different configurations.

\begin{acknowledgments}

\noindent
We acknowledge the financial support of the Academy 
of Finland via the Academy Project COPLAST (Project no. 322405).
\end{acknowledgments}

%

\section*{References} 
\bibliography{aipsamp}
\end{document}